\documentclass[twocolumn,preprintnumbers,amsmath,amssymb]{revtex4}

\usepackage{graphicx}
\usepackage{dcolumn}
\usepackage{bm}

\begin{document}
\newcommand{\kB}{k_{\mathrm{B}}}
\newcommand{\rab}{r_{1 2}}
\newcommand{\me}{\mathrm{e}}
\newcommand{\K}{\mathrm{K}}
\newcommand{\mt}{\mathrm{t}}
\newcommand{\mg}{\mathrm{g}}
\newcommand{\diff}{\mathrm{d}}
\newcommand{\pc}{p(x_0,\tau)}
\newcommand{\Ppcr}{P[p,r]}
\newcommand{\xtau}{x_\tau}
\newcommand{\nvap}{N_{\mathrm{vap}}}
\newcommand{\rcyl}{\rho_{\mathrm{cyl}}}

\title{The Role of Solvent Fluctuations in Hydrophobic Assembly}

\author{Adam P. Willard}
 \author{David Chandler}
\affiliation{Department of Chemistry, University of California, Berkeley, California 94720}

\date{\today}

\begin{abstract}
We use a coarse grained solvent model to study the self assembly of two nano-scale hydrophobic particles in water.  We show how solvent degrees of freedom are involved in the process.  By using tools of transition path sampling, we elucidate the reaction coordinates describing the assembly.  In accord with earlier expectations, we find that fluctuations of the liquid-vapor-like interface surrounding the solutes are significant, in this case leading to the formation of a vapor tunnel between the two solute particles.  This tunnel accelerates assembly.  While considering this specific model system, the approach we use illustrates a methodology that is broadly applicable. 
 \end{abstract}

\maketitle
\section{\label{sec:introduction}Introduction}
It is widely accepted that solvent water plays an important role in processes that involve interactions between hydrophobic surfaces.  For a recent review, see Ref. \cite{DC05}.  Here, we describe the role solvent plays in the self assembly of idealized nano-scale hydrophobic monomers, specifically, two hard spheres with radii $R = 1\mathrm{nm}$.   With such solutes, the solute surface exposed to solvent is sufficiently large that the solvation free energy is dominated by the solvent-solute surface free energy \cite{DC05,DMH01}.  Accordingly, the free energy of assembly is roughly
	\begin{equation*}
 	2 \pi \gamma_\mathrm{w} (R+r_\mathrm{w})[2 (R+r_\mathrm{w}) - d],
	\end{equation*}	
where $r_\mathrm{w} \approx 0.14 \mathrm{nm}$ is the radius of a water molecule, $d$ is the distance between the centers of the two cavities in the dimerized state, and $\gamma_\mathrm{w}$ is the water-vapor surface tension.  For $R =1\mathrm{nm}$, this estimate gives $\Delta E \approx -35 \kB T$, where $k_\mathrm{B}$ is Boltzmann's constant and $T$ is temperature.  With such a large binding free energy, there is little doubt that two nano-scale hard spheres will form long-lived dimers in water.  The question we consider here is specifically how the dimer forms, what dynamical pathways lead to assembly.

  The model we use to address this issue is described in the next section.  Methods of analyzing its dynamics are discussed in Section \ref{sec:Theory}.  Results and conclusions are presented in Section \ref{sec:results}.
  
\section{\label{sec:methods}The Model}
We have adapted the model of ten Wolde and Chandler\cite{PRtW02}.  In this model, the solvent is a lattice gas through which solute particles are allowed to move continuously.  Recent work\cite{TFM07} has established that the ten Wolde-Chandler model does follow from coarse graining trajectories of an atomistic model of liquid water.  This solvent model has two parameters: a lattice grid spacing, $l = 0.21 \mathrm{nm}$, and nearest neighbor coupling constant, $\epsilon = 1.51 \kB T$.  With these parameters, the lattice gas has the same surface tension and compressibility as water at standard conditions.  The chemical potential of the lattice gas we consider is $\mu = \mu_{\mathrm{coex}} + 2.25 \times 10^{-4} \kB T$, where $\mu_{\mathrm{coex}}=-3\epsilon$ is the chemical potential of the lattice gas at liquid-vapor phase coexistence.  With this choice, the model is as close to phase coexistence as is liquid water at standard conditions.
  
The two solute particles we consider are ``ideally'' hydrophobic in the sense that their only interactions with the solvent is to exclude volume.  A solute particle with a hard core radius of $R$ will thus exclude the center of solvent molecules from occupying a spherical volume of radius $R + r_\mathrm{w}$.  Modest attractions between solute and solvent, such as those between oil and water molecules, could be added to the model but with little effect on solvent fluctuations\cite{LM07}
  
The net potential energy for the model can be viewed as a free energy that results from integrating out density fluctuations occurring on length scales smaller than the lattice spacing.  For lattice gas plus solutes we take it to be 
	\begin{equation}
	H[\lbrace n_k \rbrace ; \lbrace v_l \rbrace ] \approx - \varepsilon {\sum_{i,j}}' n_i n_j + \sum_i [-\mu + \Delta \mu_{ex}(v_i)] n_i + U(r).
	\label{eq:latham}
	\end{equation}
where $n_i = 0,1$ is the lattice-gas variable for cell $i$.  The primed summation is over nearest-neighbor pairs of lattice sites, and $U(r)$ is the pair-potential of interaction between the solutes when separated by distance $r$.  The variable $v_i$ is the volume of lattice site $i$ occupied by solute particles, and $\Delta \mu_{ex}(v_i)$ is the free energy of solvation for that volume.  It is due to coarse graining that excluded volume appears as a soft (rather than hard) constraint in the model\cite{PRtW01}.  Since solvation energy at small length scales is proportional to volume,
	\begin{equation}
	\Delta \mu_{ex}(v_i) = c v_i
	\label{eq:cdef}
	\end{equation}
where $c =0.6  \kB T/l^3$ is a constant equal to the excess chemical potential per unit volume of a small hard sphere in water\cite{PRtW01}.  For $U(r)$ we use the WCA potential\cite{JDW71},
	\begin{eqnarray}
	U(r) = 4&\epsilon [ (\sigma/r)^{12} - (\sigma/r)^6 ] + \epsilon \;\;\;\;\;\;\;\;\;\;\;\; &r< 2^{1/6} \sigma \\
 	= 0& \;\;\;\;\;\;\;\;\;\;\;\; &r \geq 2^{1/6} \sigma.
	\label{eq:wca}
	\end{eqnarray}
where $\sigma = 2R$.
  
The system evolves with a stochastic dynamics moving the solvent through its discrete configurations and moving the hydrophobic spheres continuously in space.  The solvent is moved from a time $t$ to a time $t+\delta t_l$ with a Metropolis Monte Carlo\cite{fands} sweep.  A sweep consists of $M$ moves, where $M$ is the total number of lattice sites in the system.  For each move we choose a random lattice site $i$, and attempt a change of state (i.e. $n_i = 0 \rightarrow n_i = 1$ or $n_i = 1 \rightarrow n_i = 0$) with an acceptance probability consistent with the Boltzmann weight for the energy function in Eq. \ref{eq:latham}.  The volumes $v_i$ are fixed by the positions of the solute particles during this step.  We associate the sweep time with the physical time $\delta t_l = 5.0 \times 10^{-13} \mathrm{s}$, as this value approximates the correlation time for bulk liquid density fluctuation of length scale $l$ [i.e., $\delta t_l = l^2/(4 \pi^2 D)$, where $D$ is the self diffusion constant of liquid water].  Sweeps are performed as either collections of changes in state of single lattice sites, or as changes in nearest-neighbor pairs of lattice sites to conserve net solvent occupancy.  The kinetics depend upon which procedure is followed, though the mechanism for nano-particle assembly appears unaffected.  This point is discussed further in Section \ref{sec:results}.
 
The hydrophobic particles respond to the forces
	\begin{equation}
	\vec{F}_\alpha(\lbrace \vec{r}_\xi \rbrace, \lbrace n_k \rbrace) = -\vec{\nabla}_\alpha \left [U \left ( \left \lbrace \vec{r}_\xi \right \rbrace \right ) + c \sum_i v_i n_i \right ] + \vec{f}_\alpha,
	\label{eq:force}
	\end{equation}
where the subscript $\alpha$ identifies the particular solute (i.e., $\alpha = $1 or 2) and $\vec{f}$ is the random force that is the remnant of the small length-scale density fluctuations.  Because small length-scale density fluctuations are to a good extent, Gaussian\cite{GH96}, $\vec{f}$ should be Gaussian with $\langle \vec{f} \rangle = 0$ and $\langle \vert \vec{f} \vert ^2 \rangle = 6 \kB T \gamma / \delta t_s$.  We use $\gamma = 6 \pi \eta R$, and $\eta$ is the viscosity of liquid water.  With these forces, the solute positions at time $t$ progress to those at time $t + \delta t_s$ according to  
	\begin{equation}
	\vec{r}_\alpha(t+\delta t_s) = \vec{r}_\alpha(t) + \frac{\delta t_s}{\gamma} \vec{F}_\alpha(\lbrace \vec{r}_\xi(t)\rbrace,\lbrace n_i \rbrace).
  	\label{eq:langevin}
  	\end{equation}
The value of $\delta t_s$ is set as the largest time interval for which Eq (\ref{eq:langevin}) maintains detailed balance to an acceptable extent.  This gives $\delta t_s = 2.8 \times 10^{-14} \mathrm{s}$.  Trajectories are generated by the repeated application of the above procedure.  Crucially, this evolution is reversible and preserve the equilibrium distribution consistent with the energy (\ref{eq:latham}).
  
To initiate our studies of the model, the solute particles were placed randomly inside the system and the solvent was allowed to equilibrate around fixed solute particles.  After the initial solvent equilibration the solute particles were allowed to move and trajectories were recorded.  The simulation cell was a cube with a side length of $6.8 \mathrm{nm}$  and was periodically replicated in each of the three Cartesian directions.

Free energy surfaces discussed in Section \ref{sec:results} were computed using umbrella sampling\cite{fands} with a bias potential in the relevant solvent coordinate and a potential of mean force in the relevant particle coordinates.

\section{\label{sec:Theory}Theoretical Methods}
Many of the tools we used to analyze the dynamics of assembly were borrowed from transition path sampling methods.  For a review, see Ref. \cite{PGB02}.
  As the system evolves, the state of the system at time $t$ is given by $x_t$, which denotes the collection of variables $\vec{r}_1$, $\vec{r}_2$ and $\lbrace n_i \rbrace$ at time $t$.  We define a function $h(x_t)$ to be equal to 1 if $x_t$ corresponds to a dimerized configuration, and 0 otherwise.  Given an initial configuration, $x_0$ (particle positions and solvent configuration),  we define $\pc$ as the probability that a trajectory initialized from that configuration will be in a dimerized state after an observation time $\tau$.  This probability is analogous to the commitor in transition path sampling\cite{PGB02}.  While $\tau$ should not be so large as to make $\pc$ independent of $x_0$, it should be large enough to allow for dimerization.  This time scale separation will be discussed further in Section \ref{sec:results}.  The collapse probability can be written as,
	\begin{equation}
	p(x,\tau) = \frac{\langle \delta(x - x_0) h(x_\tau)\rangle}{\langle \delta(x - x_0) \rangle}
	\end{equation}
where the pointed brackets indicate an equilibrium average over initial conditions and is equivalent to an average over many trajectories, each of length $\tau$.

The contraction of this probability for the high dimensional $x$ to that for a lower dimensional $q(x)$ is,
	\begin{equation}
	\bar{p}(q,\tau) = \frac{\langle\pc \delta(q - q(x_0)) \rangle}{\langle \pc \rangle},
	\label{eq:effectivepc}
	\end{equation}
where $q(x_0)$ is the initial value of the coordinate $q$.  The $q$'s we have in mind are those that can be good reaction coordinates.  In practice we compute $\bar{p}(q,\tau)$ by averaging $N$ trajectories over many sets of initial conditions \cite{pc_sampling},
	\begin{equation}
	\bar{p}(q,\tau) \propto \frac{1}{N} \sum_{i = 1}^N h(x^{(i)}_\tau)\delta(q - q(x_0)).
	\label{eq:effectivepc2}
	\end{equation}
where $x^{(i)}_\tau$ refers to $x$ at time $\tau$ for the $i$th trajectory.
  	\begin{figure*}[t]
	\resizebox{16.5cm}{!}{\includegraphics{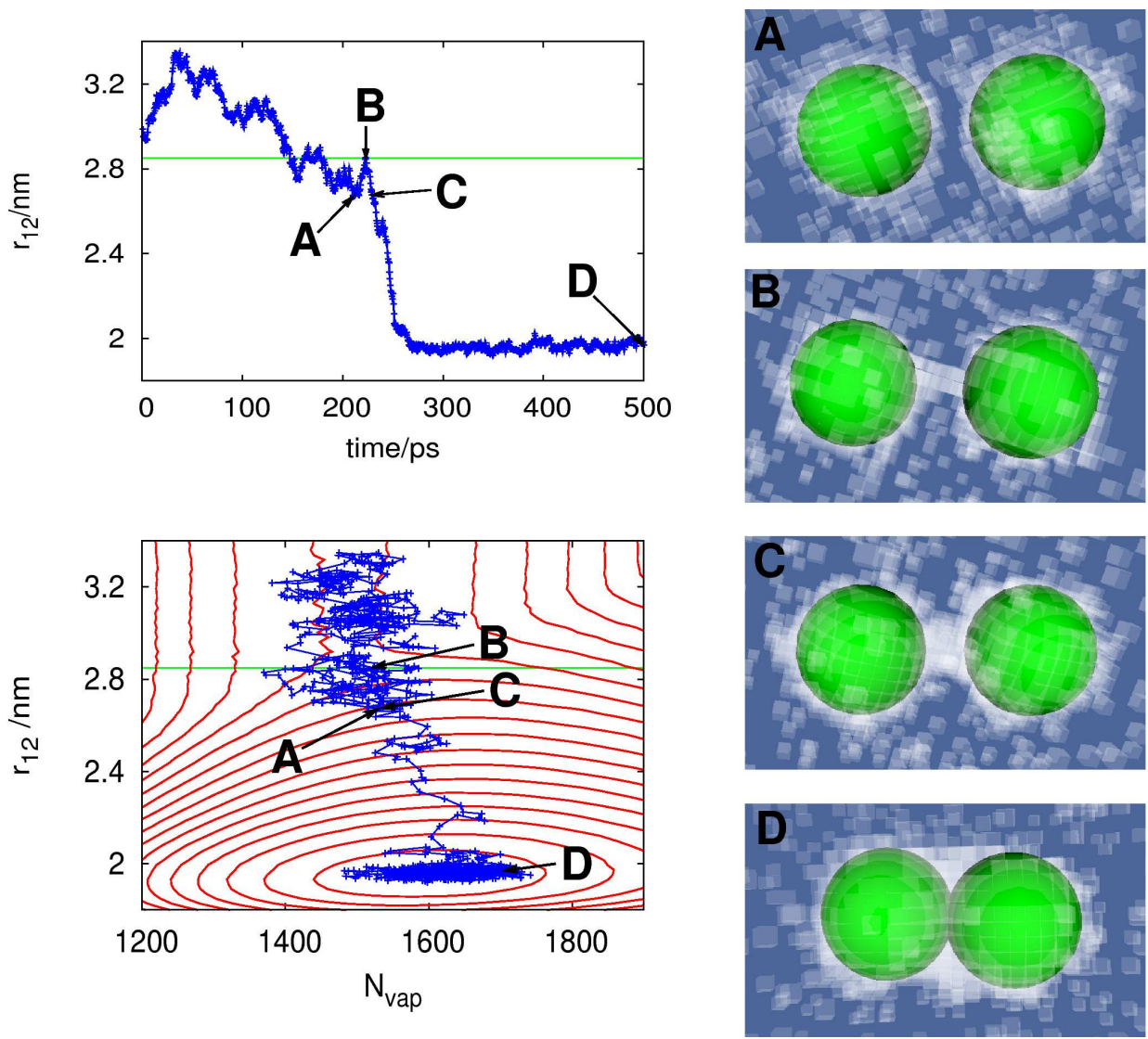}}
	\caption{Snapshots of a reactive trajectory at several points in time.  Each snapshot is labeled on the trajectory plotted in the one dimensional phase space of $\rab$ (top left) and 	the two dimensional phase space of $\rab$ and $\nvap$ (bottom left).  See text for definitions.}
  	\label{fig:trajmech}
	\end{figure*}
 
There is  an ensemble of transition state configurations, and for each member of this transition state ensemble $\pc = 1/2$.  Members of the transition state ensemble may appear diverse, but there are, however, configurational features that are common to members of the transition state ensemble which arise as signatures of the reaction mechanism.  A reaction coordinate $q$ which properly characterizes the dynamics of the reaction should also resolve those signature features common to members of the transition state ensemble.  If the coordinate $q$ does in fact resolve the important features of the reaction mechanism, members of the transition state ensemble will be narrowly distributed around $q=q^*$, with $\bar{p}(q^*,\tau) \approx 0.5$.  This expectation, plus the expectation that $p(x_0,\tau)$ is also narrowly distributed at $q^*=q(x_0)$ will provide us with a criteria to reject unsatisfactory (or poor) reaction coordinates. 
  
The distribution of $p(x_0,\tau)$ is
	\begin{equation}
	P(p;q^*) = \frac{\langle \delta( p - \pc) \delta(q^* - q(x_0)) \rangle}{ \langle \delta(q^* - q(x_0)) \rangle},
	\end{equation}
where $q(x_0)$ is the initial value of the proposed coordinate.  In practice, the distribution is computed by averaging over many trajectories with a set of $N$ initial conditions taken from an equilibrium ensemble with $q(x_0)$ constrained to $q^*$ \cite{Ppc_sampling},
	\begin{equation}
 	P(p;q^*) \propto \frac{1}{N} \sum_{i=1}^{N}\delta(p - p(x^{(i)}_0,\tau))
   	\end{equation}
\section{\label{sec:results}Results and Discussion}	
Figure \ref{fig:trajmech} shows a series of snapshots from a single reactive trajectory.
	When viewing trajectories as a function of particle separation, $\rab \equiv \vert \vec{r}_1 - \vec{r}_2 \vert$, where $\vec{r}_1$ and $\vec{r}_2$ are the positions of the two solute particles, one observes three distinct behaviors.  At large values of $\rab$, the particles exhibit independent diffusive motion.  At lower values of $\rab$, solutes are bound together in a stable dimerized state.  At intermediate values of $\rab$, the nano-particles undergo aggregation, which is manifest in trajectories as a rapid decrease in $\rab$ in time, such a rapid decrease is seen in the top left panel of figure \ref{fig:trajmech}.  The time over which this rapid decrease occurs is about 50ps in the pictured trajectory.  We have found this to be typical of all the trajectories we have studied.
  
  \subsection{\label{ssec:soluterc}Solute Reaction Coordinate}
Since $\rab$ is capable of distinguishing between the diffusive and collapsed regime, it is a natural order parameter.  To determine if $\rab$ is also a good reaction coordinate we must first define the transition state for assembly.  We define transition states as those configurations $x$ for which $p(x,\tau) = 1/2$.  This definition requires a specification the indicator function $h(x)$ and the time $\tau$.
  
\begin{figure}[hbt]
	\resizebox{\columnwidth}{!}{\includegraphics{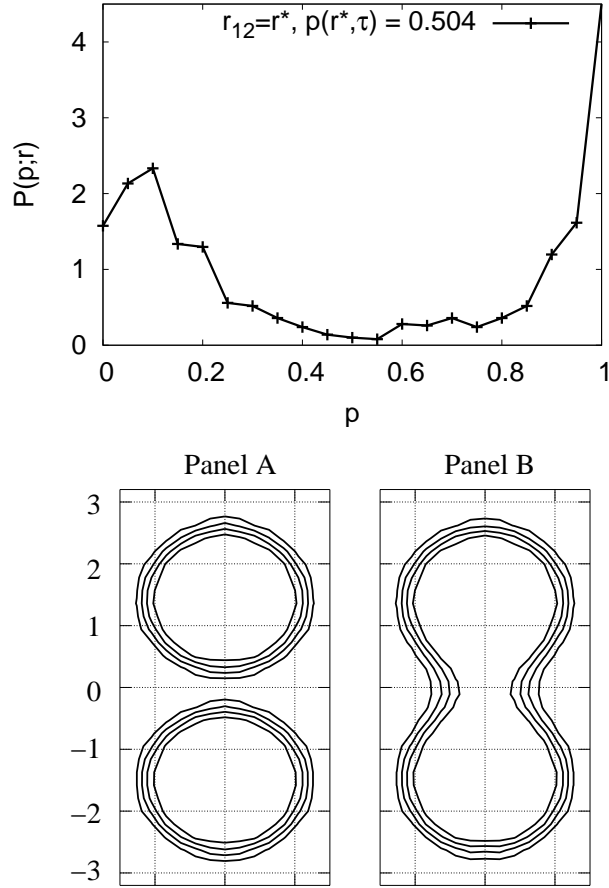}}
	
	\caption{(top)The distribution of collapse probabilities, $P(p;r)$ for $\rab=r^*=2.85 \mathrm{nm}$(lines are guides to the eye).
	(bottom)The cylindrical average of the initial solvent configuration around the solute particle at a distance $\rab = r^*$. 
	  Panel A is an average over the least reactive initial solvent configurations ($\pc \vert_ {\rab=r^*} \leq 0.10$) and panel B is an average over the most reactive initial solvent configurations ($\pc \vert_{\rab=r^*} \geq 0.90$). 
	  In each panel the center of the particles have $\langle n_i \rangle = 0$ and contour lines are drawn at increasing increments of $\langle n_i \rangle = 0.2$.}
	\label{fig:ICdenmap}
	\end{figure}

Judging from the trajectories, we have taken,
	\begin{eqnarray}
  	h(x) &=& 1 \;\;\;\; \rab \leq 2.1 \mathrm{nm} \\ 
  	&=& 0 \;\;\;\; \rab > 2.1 \mathrm{nm}.
  	\end{eqnarray}
and we take the observation time $\tau = [(\rab - 2.1 \mathrm{nm})/0.6 \mathrm{nm}]50 \mathrm{ps}$.  The solid horizontal line drawn in figure \ref{fig:trajmech} represents the value of $\rab = r^* = 28.5\mathrm{\AA}$ for which $\bar{p}(r^*,\tau) \approx 0.5$.  The top panel of Fig. \ref{fig:ICdenmap} shows $P(p;r)$ for $\rab = r^* = 2.85 \mathrm{nm}$.
We see that $P(p;r)$ is bimodal, indicating that a given configuration drawn from an equilibrium ensemble at $\rab = r^*$ is typically reactive (resulting trajectories mostly result in aggregation) or nonreactive, but not a reactive threshhold.
We conclude that $\rab$ alone does not suffice in describing the mechanism of hydrophobic assembly.
   
\subsection{\label{ssec:solventrc}Solute and Sovent Reaction Coordinate}
Since the solute particles are spherically symmetric, and the solvent is isotropic, the only information contained in the initial conditions which is not contained in $\rab$ involves the solvent configuration.  The bottom panels of Fig. \ref{fig:ICdenmap} show the average initial solvent density between the two solute particles at a separation of $r^*$.  The least reactive initial solvent configurations have a high solvent density in the volume between the two particles while the most reactive initial solvent configurations have a vapor tunnel which connects the solvent cavities of the two solute particles.  This is reminicent of the vapor tunnel which forms as a precursor to the drying of a metastable liquid confined between extended hydrophobic surfaces \cite{KL98,AL00,XH05}.

The difference between panels A and B in Fig. \ref{fig:ICdenmap} indicate that a solvent coordinate should be capable of resolving a vapor tunnel.  A reaction coordinate which differentiates between panels A and B and captures the interfacial fluctuations is the total number of vapor-like ($n_i = 0$) lattice sites in the vicinity of the hydrophobic solute particles, $\nvap$, which we define as,
  	\begin{equation}
  	\nvap = \sum_i(1-n_i)\Theta(r_\mathrm{c} - \mathrm{min}(\vert \vec{r}_i - \vec{r}_1\vert,\vert \vec{r}_i - \vec{r}_2 \vert))
  	\label{eq:nvap}
  	\end{equation}
where $\Theta(x)$ is the Heaviside function (i.e., 1 for $x > 0$ and 0 for $x < 0$) and $r_\mathrm{c}$ is a cutoff radius, which we have taken to be four lattice spacings beyond the surface of a hydrophobic sphere, see Fig. \ref{fig:OP3}, in particular, $r_\mathrm{c} = 1.84 \mathrm{nm}$.
  
The bottom left panel of Fig. \ref{fig:trajmech} shows a reactive trajectory plotted in the two dimensional reaction coordinate of $\rab$ and $\nvap$ along with contours of the free energy surface.  Focusing first on the contours of the free energy surface in Fig. \ref{fig:trajmech} we observe that the diffusive regime is shaped like a trough which is parabolic in the direction of $\nvap$ near the mean (fluctuations in $\nvap$ about the mean are Gaussian).  The diffusive trough in the free energy surface funnels into the deep basin associated with the aggregated dimer state.  The difference in free energy between the minimum of the diffusive basin and the bottom of the aggregated basin is about $64 \kB T$.  Near the transition state, fluctuations in $\nvap$ are non-Gaussian, and large fluctuations, especially towards larger values of $\nvap$ are stabilized.  In addiction, the assembly of the hydrophobic nano-particles is an activationless process, as there is no appreciable barrier in the free energy surface.  As a result, when the particles are separated by distance $r^* = 2.85 \mathrm{nm}$ the solvent fluctuations required to create a vapor tunnel are of the order $\kB T$, the thermal energy of the system.
  
The collapse probability distribution $P(p;r,N)$ was computed for two points in the transition state region.  We found that just as for  the one-dimensional $\rab$ (plotted in in Fig. \ref{fig:ICdenmap}), the distribution of collapse probabilities are bimodal for both points in the transition state region for the two dimensional coordinate.  This is a sign that a solvent coordinate with greater spatial resolution is required to characterize this transition.  Although $\nvap$ is capable of distinguishing between panels A and B in Fig. \ref{fig:ICdenmap}, it does not reliably resolve a vapor tunnel in individual trajectories.  This can be seen by comparing snapshots A and C in Fig. \ref{fig:trajmech}, which appear very different in their solvent configurations, but lie in almost identical regions of the two-dimensional phase space $(\rab,\nvap)$.
  
 \subsection{\label{ssec:resolution}Further Resolution of Sovent Dynamics}
\begin{figure}[hbt]
	\resizebox{\columnwidth}{!}{\includegraphics{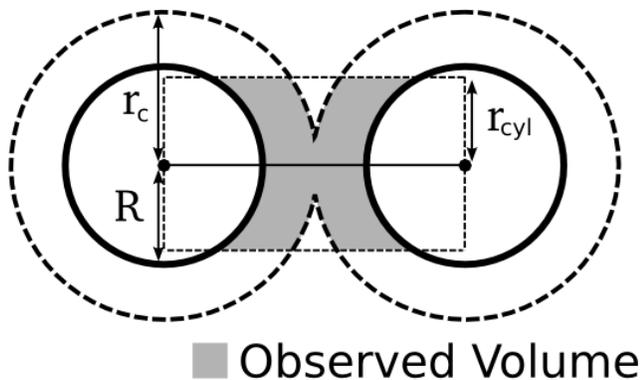}}
	\caption{A two dimensional depiction of the volume (in grey) over which the solvent coordinate $\rcyl$ is computed.}
	\label{fig:OP3}
	\end{figure}

To achieve greater spatial resolution we restrict our observation to the volume between the solute particles, where the formation of the vapor tunnel occurs (see shaded region in Fig. \ref{fig:OP3}).  
The solvent density in this region, $\rcyl$ is
   	\begin{equation}	
	\rho_\mathrm{cyl} = \frac{\sum_i n_i V_i(\vec{r}_1,\vec{r}_2)}{\sum_i V_i(\vec{r}_1,\vec{r}_2)}
	\label{eq:OP3}
	\end{equation}
where 
	\begin{eqnarray*}
	V_i(\vec{r}_1,\vec{r}_2) \;\; = &&\Theta(r_\mathrm{c} -  \mathrm{min}(\vert \vec{r}_i - \vec{r}_1\vert,\vert \vec{r}_i - \vec{r}_2 \vert)) \\
	&&\times \Theta(\mathrm{min}(\vert \vec{r}_i - \vec{r}_1\vert,\vert \vec{r}_i - \vec{r}_2 \vert) - R) \\
	&&\times \Theta(r_\mathrm{cyl} - \vert (\vec{r}_{i} - \vec{r}_1) \times \vec{r}_{12}/\rab \vert) \\
	&&\times \Theta((\vec{r}_i - \vec{r}_2) \cdotp \vec{r}_{12})  \Theta ((\vec{r}_1 - \vec{r}_i) \cdotp \vec{r}_{12})
	\end{eqnarray*}
with $r_\mathrm{cyl} = 0.7 \mathrm{nm}$.

\begin{figure}[hbt]
	\resizebox{\columnwidth}{!}{\includegraphics{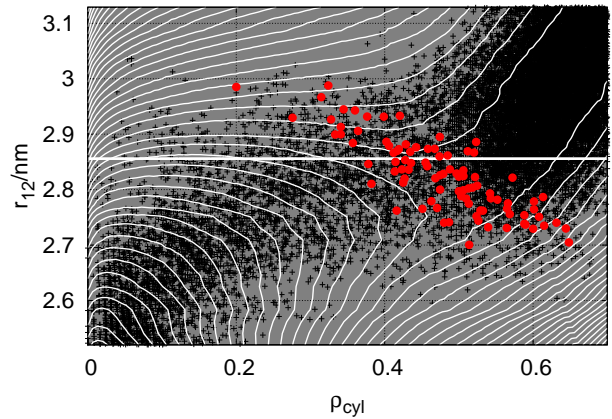}}
	\caption{Points (black) visited by 100 trajectories projected onto the $\rab$ and $\rcyl$ plane, with contours of the free energy surface overlaid (white lines).  All trajectories were generated with initial values of $\rab \ge  2.8 \mathrm{nm}$. Contour lines are in increments of $1\kB T$.  Points shown in red are members of the transition state ensemble. }
	\label{fig:OP3FE}
	\end{figure}

Figure \ref{fig:OP3FE} plots points visited by many reactive trajectories along with the contour lines of the free energy surface for $(\rab,\rcyl)$.
With the spatial resolution of $\rcyl$ we observe a pronounced depletion of the solvent density between the solutes prior to the passing through of the transition bottleneck.  The red points in figure \ref{fig:OP3FE} are members of the transition state ensemble, which are clustered near the transition bottleneck of the free energy surface.

\subsection{\label{ssec:mech}Mechanism of Dimerization}
This figure, along with Fig. \ref{fig:ICdenmap} provides us with a description of the aggregation.  The transition begins when the particles have diffused to positions that bring their surrounding solute cavities to within the range accessible by equilibrium fluctuation of the cavity interface.  As the interfaces surrounding each solute particle samples many configurations the cavities come into contact which creates the beginning of a vapor tunnel.  The growth of the vapor tunnel is manifest on Fig. \ref{fig:OP3FE} as the rapid decrease of the density, $\rcyl$, between the solutes prior to passing through the bottle neck in the free energy surface.  The existence of a vapor tunnel leads to an unbalancing of the solvent induced force on the solute particles.  This unbalanced force arises because the solute particles are being pushed by solvent in every direction accept the direction of the vapor tunnel.  The result is a net solvent induced force on each solute in the direction of the vapor tunnel.  The particles are pushed closer together and are eventually squeezed into physical contact by the solvent.
  
\subsection{\label{ssec:masscons}Role of Solvent Mass Conservation}
Solvent dynamics were carried out along a Markov chain, the details of which are provided in Section \ref{sec:methods}.  One consequence of using single lattice site moves, is that the net solvent density is allowed to fluctuate in a fashion that does not account for solvent mass conservation.  To examine the extent to which this conservation is important, we have generated trajectories with Kawasaki\cite{KK66} neighbor exchange dynamics.  These dynamics are carried out just like the Metropolis dynamics described in Section \ref{sec:methods}, but instead of attempting changes of state for single lattice sites, the states of two neighboring lattice sites are exchanged.  The acceptance probability, of course, is consistent with the Boltzmann weight for the energy in Eq. \ref{eq:latham}.  The free energy surface, an equilibrium property, is invariant to the choice of dynamics.  Those aspects of trajectories determined by statistics are unchanged.  Specifically, under both the mass-conserving and non mass-conserving dynamics, dimerization is preempted by the formation of a vapor tunnel which accelerates assembly.  In both cases the onset of accelerated assembly occurs near $\rab=r^*=2.85 \mathrm{nm}$.  But the rate at which solutes are pushed together in this region are very different for the two forms of solvent dynamics.  Specifically, after passing through the transition state by forming a stable vapor tunnel, trajectories with the mass-conserving solvent dynamics take about 2000ps to dimerize, more than one order of magnitude slower than the dynamics without mass conservation.  Thus, although the mechanism of hydrophobic assembly is invariant to solvent mass conservation, the rate of passing through the transition state ensemble does depend upon conservation, so that prefactors to an Arrhenius rate constant expression depend upon it significantly.

\subsection{\label{ssec:attract}Role of Solute-Solvent Attractions}
The model considered in detail here contains idealized hydrophobic solutes who's only interactions with the solvent is to excluded it from occupying a specific volume.  In real physical systems, there are additional solute-solvent interactions, such as Van der Waals attractions which can also play a role in the dynamics.  Recent work\cite{LM07} has described how the average solvent density surrounding hydrophobic solutes responds linearly to attractive solute-solvent forces.  This response increases the mean density, and shifts the position of the surrounding interface closer to the solute surface.  To the extent that solvent fluctuations are Gaussian, the nature of the interfacial fluctuations are unchanged.  Thus, for fixed solute positions, the addition of such attractions necessarily increases the distance between the interfaces surrounding  different solute particles.  As a result, larger solvent fluctuations are required to form a vapor tunnel, leading to an increase in the activation energy to reach the transition state.  A free energy barrier to the formation of a vapor tunnel that is more than a few $\kB T$, can be reduced (and even avoided all together) if the solutes diffuse closer together.  Thus, attractive forces that do not destroy the interface by pinning it to the solute surface (in which case the solute could hardly be considered hydrophobic) do not qualitatively change the mechanism discussed here.  The effect of such attractions would, however, cause a decrease in the value of the distance $\rab = r^*$, the distance at which the onset of dimerization is observed, in effect, small attractive forces make the solvated particles look smaller. 
  
This work was supported in its initial stages by the Director, Office of Science, Office of Basic Energy Sciences, Chemical Sciences, Geosciences, and Biosciences Division, U.S. Department of Energy under Contract No. DE-AC02-05CH11231, and in its final stages by the National Institutes of Health.

\bibliographystyle{unsrt}

\begin{thebibliography}{1}
\bibitem{pc_sampling}  In this study $\bar{p}(q,\tau)$ for the one-dimensional coordinate was computed by taking $N$ to be the product of 50 individual trajectories for each of 1000 unique configurations. For the two dimensional coordinate $N$ was the product of 100 individual trajectories for each of 50 unique initial configurations.
\bibitem{Ppc_sampling} In this study, for the one-dimensional coordinate, $P(p;r_{12})$ was computed by taking $N=1000$ initial configurations each averaged over 100 individual trajectories. For the two-dimensional coordinate, $P(p;r_{12},N_{\mathrm{vap}})$ was computed by taking $N=500$ initial configurations each averaged over 100 individual trajectories.
\end{thebibliography}

\end{document}